\documentclass[12pt,a4paper,final]{iopart}
\usepackage{iopams}
\usepackage{graphicx,lipsum}
  \expandafter\let\csname equation*\endcsname\relax
  \expandafter\let\csname endequation*\endcsname\relax
\usepackage{amsmath,color}

\usepackage{pgfplots}

\usepackage{float}
\usepackage{physics}
\def\gappeq{\mathrel{ \rlap{\raise.5ex\hbox{$>$}}
                      {\lower.5ex\hbox{$\sim$}} } }
\def\lappeq{\mathrel{ \rlap{\raise.5ex\hbox{$<$}}
                      {\lower.5ex\hbox{$\sim$}} } }
\usepackage{bm}
\usepackage{color}
\usepackage{epstopdf}

\newcommand{\nick}[1]{\textcolor{black}{#1}}
\newcommand{\tom}[1]{\textcolor{black}{#1}}

\begin{document}

\title{Persistent current formation in double-ring geometries}

\author{T. Bland}
\address{Joint Quantum Centre Durham--Newcastle, School of Mathematics, Statistics and Physics, Newcastle University, Newcastle upon Tyne, NE1 7RU, United Kingdom}
\author{Q. Marolleau}
\address{Joint Quantum Centre Durham--Newcastle, School of Mathematics, Statistics and Physics, Newcastle University, Newcastle upon Tyne, NE1 7RU, United Kingdom}
\address{Laboratoire Charles Fabry, Institut d'Optique Graduate School, CNRS, Universit\'e Paris-Saclay, 91120 Palaiseau, France}
\author{P. Comaron}
\address{Joint Quantum Centre Durham--Newcastle, School of Mathematics, Statistics and Physics, Newcastle University, Newcastle upon Tyne, NE1 7RU, United Kingdom \\
Institute of Physics Polish Academy of Sciences, Al.~Lotnik\'ow 32/46, 02-668 Warsaw, Poland}
\author{B. A. Malomed}
\address{Department of Physical Electronics, School of Electrical
Engineering, Faculty of Engineering, Tel Aviv University, Tel Aviv 69978,
Israel}
\author{N. P. Proukakis}
\address{Joint Quantum Centre Durham--Newcastle, School of Mathematics, Statistics and Physics, Newcastle University, Newcastle upon Tyne, NE1 7RU, United Kingdom}

\begin{abstract}
\noindent Quenching an ultracold bosonic gas in a ring across the
Bose-Einstein condensation phase transition is known, and has been experimentally observed, to lead to the spontaneous emergence of persistent currents. The present work examines how these phenomena generalize to a system of two experimentally accessible \nick{explicitly two-dimensional} co-planar rings with a common interface, or to the related lemniscate geometry, and demonstrates an emerging independence of winding numbers across the rings, which can exhibit flow both in the same and in opposite directions. The \nick{observed} persistence of such findings in the presence of dissipative coupled evolution \nick{due} to the local character of the domain formation across the phase transition and topological protection of the randomly emerging winding numbers should be within current experimental reach. 
\end{abstract}

\maketitle

\section{Introduction}


At the critical point of a second order phase transition, the symmetry of a
system is spontaneously broken, with both relaxation time and correlation
length diverging \cite{zurek1985cosmological}. In the transition region, the
resulting state is not perfectly ordered, rather building a mosaic pattern
of frozen coherent domains, with defects spontaneously emerging at their
boundaries, thus becoming embedded within the system's growing coherent
global state. The study of emerging defect dynamics and their subsequent
relaxation has been a topic of active research over the course of many years.

This effect was first discussed by Kibble in a cosmological context, setting
an upper bound on the domain regions \cite{kibble1976topology}.
Zurek extended this by quantifying the emerging domain size on the basis of
the universality of critical slowing down \cite{zurek1985cosmological}.
Consideration of the role of the quench timescale in finite-duration and linear quenches led to the universal Kibble-Zurek mechanism, which has been
observed in a variety of complex systems \cite{kibble2007phase}, including
liquid crystals \cite{chuang1991cosmology}, liquid helium \cite%
{ruutu1996vortex,bauerle1996laboratory}, superconducting loops \cite%
{carmi2000observation,monaco2002zurek,monaco2003spontaneous,monaco2009spontaneous}%
, ion chains \cite{ulm2013observation}, Bose-Einstein condensates (BECs)
\cite{sadler2006spontaneous,weiler2008spontaneous,lamporesi2013spontaneous,%
donadello2016creation,navon2015critical,corman2014quench,chomaz2015emergence,%
braun2015emergence,liu2018dynamical,ko2019kibble}, and, recently, in Rydberg lattices \cite{keesling2019quantum}. In ring
geometries, like the original configuration considered by Zurek \cite%
{zurek1985cosmological}, the frozen phase of the wave function may lead to
emergence of a supercurrent of integer topological charge $q$, i.e.~a $2\pi q
$ phase winding along the ring. The focus of numerous previous studies has
been to show how the supercurrent charge scales with the quench time through
the phase transition \cite%
{monaco2002zurek,monaco2003spontaneous,monaco2009spontaneous,corman2014quench,das2012winding,aidelsburger2017relaxation}%
, thus verifying the applicability of the Kibble-Zurek scaling law.
\nick{In particular, the numerical work of Ref.~\cite{das2012winding} provided a detailed visualization of, and highlighted, the role of local phase formation and subsequent evolution to a stable persistent current in the context of a one-dimensional model.}
\nick{ Although such Kibble-Zurek scaling only applies to finite duration quenches,
the
underlying phenomenon of spontaneous symmetry breaking and generation of
defects are at the heart of any crossing through a second-order phase
transition, 
also
including the numerically simpler instantaneous quenches,
which } provide crucial information for the formation dynamics of
coherence in macroscopic systems and on the critical universal properties of the system.

Dynamical quenches (whether instantaneous or gradual) are in fact a critical
ingredient of envisaged circuits in the emerging field of atomtronics, a
highly-promising interdisciplinary field at the interface between
matter-wave optics and the photonics/semiconductor technologies \cite%
{seaman2007atomtronics,schlosser2011scalable}, which has also been suggested
as a potential platform for quantum-information devices, such as qubits \cite%
{aghamalyan2016atomtronic}. The primary goal of atomtronics is to use the
precise control, admitted by the ultracold quantum matter, to generate
highly coherent circuits of neutral atoms, which are analogous to
conventional solid-state systems, but with the potential to offer much
improved and/or otherwise inaccessible technological applications \cite%
{amico2005quantum,amico2017focus}. Available atomtronic circuits include, in
particular, variable-resistance RLC schemes \cite%
{li2016superfluid,eckel2016contact,gauthier2019quantitative},
Josephson-junction SQUIDs \cite%
{ramanathan2011superflow,ryu2013experimental,eckel2014hysteresis,mathey2016realising}%
, and diodes \cite{labouvie2015negative}.

In this work, we explore the formation and structure of spontaneously
generated supercurrents induced by an \emph{instantaneous} crossing of the
phase transition to the BEC state in a co-planar, side-by-side, double-ring
geometry. This set-up has been chosen as a cold-atom analogue of a qubit
made from adjacent superconducting loops, known as the Mooij-Harmans qubit
\cite{mooij2005phase,mooij2006superconducting,astafiev2012coherent}. Its
operation relies on the use of coherent quantum-phase slips to coherently
transport vortices through Josephson links. A theoretical proposal to
implement this phenomenon in two-component quantum gases has been made \cite%
{gallemi2015coherent}, with Rabi coupling driving the phase slips
\nick{with experimentally accessible spinor two-component gases~\cite{beattie2013persistent} constituting a potential platform for such work.}
Gaining control over the tunnelling of persistent currents in a double-ring geometry
would closer emulate the geometry of the original theoretical scheme.
Recently, several proposals for two parallel/stacked rings have shown that
the winding number can tunnel between rings due to the creation of fluxons
at their boundary \cite{richaud2017quantum,oliinyk2019tunneling}, \tom{and at the single-particle level persistent current tunnelling has been predicted between arrays of adjacent rings and in similar configurations \cite%
{polo2016geometrically,pelegri2017single,%
pelegri2019topological,pelegri2019topological2}.}

The aim of this work is to explore the \nick{full two-dimensional (2D)} formation dynamics of spontaneous
supercurrents in such a side-by-side geometry, and demonstrate the resulting
(perhaps somewhat counter-intuitive) independence of the two connected
rings, despite their density overlap and coupled dynamics.
After presenting our model,
in the form of the \nick{commonly used} stochastic projected Gross-Pitaevskii equation (SPGPE) in
Sec.~\ref{sec:method},
we analyze the formation of persistent currents in
the double-ring trap geometry (Sec.~\ref{sec:drr}), and discuss the
distribution of observable persistent-current states. We then explicitly
demonstrate how
the double-ring setting can be reduced to that for two independent \nick{2D} annuli
(Sec.~\ref{sec:srr}): \nick{doing so, we are able to draw analogies to Zurek's arguments, and also demonstrate the extension of previous 1D SPGPE single-ring simulations to an explicitly 2D setting}. Our
findings \nick{for the double-ring structures} are shown to be robustly insensitive to details of the geometry, by
explicitly verifying the findings in the lemniscate (figure-of-eight)
configurations in Sec.~\ref{sec:trap}.
Identifying regimes which are optimal for experimental observation paves the
way for the use of such geometries for the design of potential qubit
operations.

\section{The theoretical model}

\label{sec:method}

Quench dynamics in quantum gases are well modelled by the above-mentioned
stochastic (projected) GPE, which provides a numerically tractable effective field theory for low-lying `coherent', or `classical' modes of the system. First proposed in
\cite{Stoof01} to study the experimentally-observed reversible crossing of
the BEC phase transition \cite{Stamper-Kurn98}, it was developed
independently with the addition a projection procedure in order to separate
`coherent' and `incoherent' constituents of the dynamics \cite%
{Gardiner_2003,Bradley_2008,blakie2008dynamics,proukakis2008finite}. This
approach has become the workhorse for modeling the condensate formation across
different platforms \cite%
{weiler2008spontaneous,liu2018dynamical,das2012winding,Proukakis13,Proukakis03,%
Proukakis_Schmiedmayer_2006,Proukakis09,Zurek_2009,%
Damski_Zurek_2010,Cockburn_2012,rooney2013persistent,%
Su13,De_2014,Liu_2016,gallucci2016engineering,%
Kobayashi_16a,Kobayashi_16b,Eckel_2018,%
Ota_2018,comaron2019quench}. %
\nick{These include a number of simulations explicitly performed in a ring trap geometry~\cite{das2012winding,rooney2013persistent,gallucci2016engineering}, of direct relevance to the present analysis. Specifically, 1D SPGPE simulations by Das {\em et al.}~\cite{das2012winding}, under controlled finite-duration quenches to condensation from noisy initial conditions, revealed very clearly the local character of the formation of phase, and its long-term dynamics leading to the spontaneous formation of persistent currents (in agreement with Kibble-Zurek). The same model was used to study spontaneous Josephson vortex formation across two linearly-coupled 1D ring traps \cite{Su13}.
In 2D rings, vortex decay and persistent-current
formation were considered in the presence of external stirring \cite{rooney2013persistent}, closely matching experimental
data \cite{neely2013characteristics}, while spontaneous emergence of persistent currents in 2D ring traps was discussed in Ref.~\cite{gallucci2016engineering}, as a limiting factor in the study of controlled generation and stability of counter-propagating dark soliton pairs.
As the SPGPE models the dynamics of the condensate and low-lying modes of the classical field $\psi(\textbf{r},t)$, it does not include purely  quantum effects like entanglement between persistent current states, which should be kept in mind when considering future atomtronic applications.}

In our current implementation, building on our earlier works \cite%
{liu2018dynamical,Liu_2016,Ota_2018,comaron2019quench}, individual trajectories in the coherent
sector of the dynamics are governed by the stochastic equation of motion
\cite{proukakis2008finite},
\begin{equation}
i\hbar \frac{\partial }{\partial t}\psi (\mathbf{r},t)=\hat{\mathcal{P}}%
\Bigg\{(1-i\gamma )\bigg[\hat{\mathcal{H}}_{\text{GP}}-\mu \bigg]\psi (%
\mathbf{r},t)+\eta (\mathbf{r},t)\Bigg\}\,,  \label{eqn:SPGPE}
\end{equation}%
describing their coupling to the incoherent sector, where
\begin{equation}
\hat{\mathcal{H}}_{\text{GP}}=-\frac{\hbar ^{2}}{2m}\nabla ^{2}+V_{\text{ext}%
}(\mathbf{r})+g_{\text{2D}}|\psi (\mathbf{r},t)|^{2}\,,
\end{equation}%
is the Gross-Pitaevskii operator, \nick{and  $m$ is the atomic mass of $^{87}\mathrm{Rb}$ atoms}. Here, {${g_{\text{2D}}=\sqrt{8\pi }\hbar
^{2}a_{s}/ml_z}$} is the two-body interaction strength in the
two-dimensional (2D) geometry, determined by the \textit{s}-wave interaction
strength $a_{s}$, $\mu $ is the chemical potential, and $l_z=\sqrt{%
\hbar /m\omega _z}$ is the transverse confinement scale imposed by
the harmonic-oscillator trap $V(z)=m\omega_{z}^{2}z^{2}/2$.
\nick{In our explicitly 2D study, we adopt tight transverse confinement $\omega _{z}=2\pi \times 1000$ Hz, in order to explicitly satisfy the quasi-2D condition
 $\hbar\omega_z\gg\mu$.}
The complex Gaussian noise is characterized by
correlations ${\expval{\eta(\textbf{r},t)\eta(\textbf{r}',t)}=0}$ and ${\expval{\eta^\star(\textbf{r},t)\eta(\textbf{r}',t')}=2\gamma\hbar  k_{B}T
\delta (\mathbf{r}-\mathbf{r}^{\prime })\delta (t-t^{\prime })}$, where $T$
is the bath temperature, $\gamma$ is the growth rate, and
the asterisk stands for the complex conjugate.
In this work, we fix $\gamma =0.05$, and conclude via numerical testing that our steady-state
results are insensitive to the exact value of $\gamma $, \nick{a statement which we have explicitly numerically confirmed here in the range $0.001\leq \gamma \leq \,0.2$, which constitute values consistent with both theoretical expectations \cite{blakie2008dynamics,proukakis2008finite} and previous successful comparisons to experiments \cite{liu2018dynamical,rooney2013persistent,Ota_2018,comaron2019quench}.
}
Projector $\hat{\mathcal{%
P}}$ implements the energy cut-off,
ensuring that the occupation of the largest included mode has average
occupation of order unity. The energy cut-off is adopted as 
$\epsilon _{\text{cut}}(\mu ,T)=k_{B}T\log (2)+\mu \,$ \tom{--derived by setting the Bose-Einstein distrubtion $f(\epsilon_\text{cut})=1$ and solving for $\epsilon_\text{cut}$ \cite{rooney2010decay}--} 
setting the spatial numerical grid for simulations with spacing $\Delta
x\leq \pi /\sqrt{8m\epsilon _{\text{cut}}}$ \cite{blakie2008dynamics}. The
kinetic energy for a winding number $n_{w}$ around a ring of radius $R$ is
given by $\hbar ^{2}n_{w}^{2}/2mR^{2}$, thus for the parameters we consider
here the adopted energy cut-off limits the winding numbers to $%
|n_{w}|\lesssim 80$, far beyond the range we address in this work.

Focusing on experimentally relevant geometries for $^{87}\mathrm{Rb}$ atoms (%
${m=1.443\times 10^{-25}}$kg and ${\tilde{g}\equiv mg_{\text{2D}}/\hbar
^{2}=0.077}$), we fix the chemical potential ${\mu =25k_{\mathrm{B}}~}%
\mathrm{nK}$ and temperature $T=10$ nK, chosen to be much lower than the
critical temperature of the Berezinskii-Kosterlitz-Thouless (BKT)\
transition (in the thermodynamic limit) \cite{prokofev2001critical},
\begin{equation}
T_{\text{BKT}}^{\infty }=\frac{\pi \mu }{k_{B}\tilde{g}\log \left( C/\tilde{g%
}\right) }\,,
\label{eqn:tbkt}
\end{equation}%
where $C\sim 13.2\pm 0.4$ \cite{prokofev2002two}. In our parameter range,
this amounts to ${T/T_{\text{BKT}}^{\infty }\approx 0.05}$, i.e.~an
essentially low-temperature setting. 

To simulate the quench-induced dynamics, we simulated Eq.~\eqref{eqn:SPGPE} \tom{using the software package XMDS \cite{dennis2013xmds2}}, starting from initial condition $\psi
(x,y)=0$, with a random realization of the initial noisy field $\eta
(x,y)$, which is devoid of any phase coherence. This is akin to an input
condition with $N$ atoms, all with the initial energy larger than the
cut-off energy $\epsilon _{\text{cut}}$, which enter the cut-off region at a
rate governed by $\gamma$ until
the gas reaches thermal equilibrium (actual values of $N$ are given below).
This can be seen as an instantaneous thermal quench (cooling) from $T\gg T^\infty_{\text{BKT}}$ to $T=10~$nK$\ll T^\infty_{\text{BKT}}$.
\nick{To confirm the broad validity of our findings, we have also explicitly verified that temperature quenches from pre-formed equilibrated 2D thermal clouds at temperatures $T > T_{\rm BKT}^{\infty}$ (rather than from an initial noisy condition) yield the same qualitative findings and practically identical long-term winding number combinations (histograms; see subsequent Fig.~\ref{fig:drhist}(b)).}


\section{Instantaneous quench in the double-ring geometry}

\label{sec:drr}

\begin{figure}[tbp]
\centering
(a)\\\includegraphics[width=0.38\columnwidth]{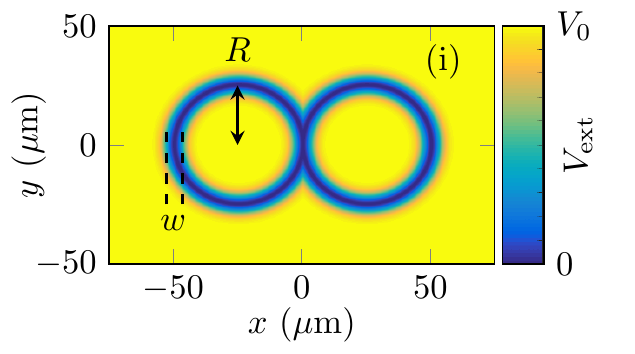}
\includegraphics[width=0.29\columnwidth]{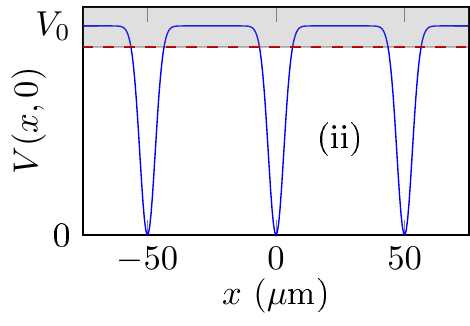}
\includegraphics[width=0.31\columnwidth]{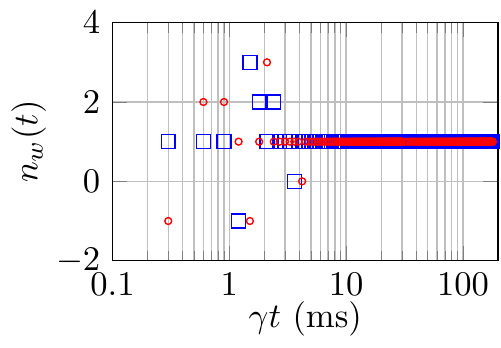}\\
(b)
\includegraphics[width=1\columnwidth]{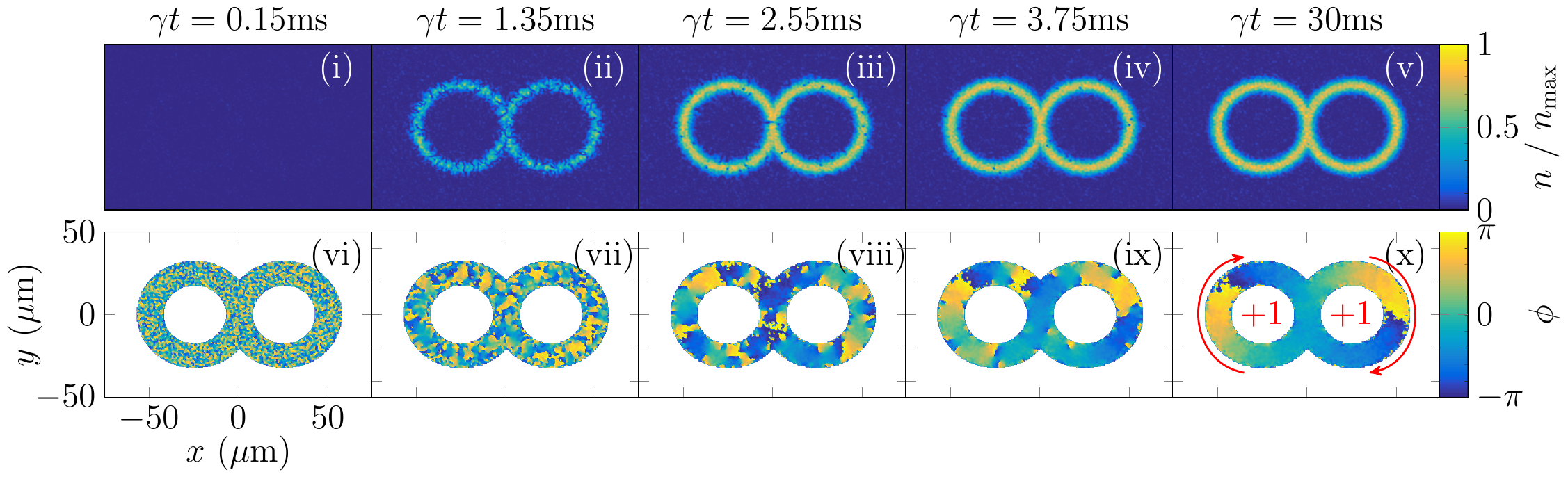}\\
\caption{Condensate growth in the double-ring geometry, with ring radius ${R=25~\mathrm{\protect\mu }}$m, width ${w=6~\mathrm{\protect\mu }}$m, and the
inter-ring shift is $\protect\delta =0$, see Eq.~(\protect\ref{delta}). (a),~(i): The double-ring potential, (ii) its cross section along $y=0$ (blue solid curve) and (iii) the evolution of the winding number over scaled time $\gamma t$ for the left (blue squares) and right (red circles) rings. (b), (i)-(v): The density evolution
in scaled time, with $n_{\text{max}}=73\mathrm{\protect%
\mu }$m$^{-2}$. (vi)-(x): The corresponding phase profiles. For clarity, these have been masked by the Heaviside function $\Theta (0.9V_{0}-V_{\text{dr}}(x,y))$, which filter out the contributions falling within the gray shaded area in (a)(ii).}
\label{fig:dr}
\end{figure}

We consider the in-plane side-by-side double-ring (dr) configuration shown
in Fig.~\ref{fig:dr}(a). \tom{This structure is also known as the 2-torus}. It is defined by the 2D potential %
%
\begin{align}
V_{\text{dr}}(x,y) =V_{0}\,\text{min}\Big(&1-\exp \left[ -2\left( \rho
(x-x_{0},y)-R\right) ^{2}/w^{2}\right] ,  \notag \\
& \,\,1-\exp \left[ -2\left( \rho (x+x_{0},y)-R\right) ^{2}/w^{2}\right] %
\Big)\,,  \label{eqn:dbpot}
\end{align}%
where $\rho (x,y)=\sqrt{x^{2}+y^{2}}$, and the centres of the two rings with radius $R
$ and width $w$ are set at
\begin{equation}
y=0,x=\pm x_{0}\equiv \pm \left( R+\delta \right) ,  \label{delta}
\end{equation}
$\delta \geq 0$ being a shift that separates the two side-by-side rings. The choice $R>w\gg\xi$, for healing length $\xi=\hbar/\sqrt{m\mu}$, gives the rings 2D character and thus justifies the use of $T^\infty_\text{BKT}$ as a measure of critical temperature.
Further, the height of the potential, $V_{0}=27.5{k_{\mathrm{B}}~}\mathrm{nK}%
\,>\mu $, is fixed throughout the work. This potential is the double-ring
extension of the single-ring trap used in Ref.~\cite{murray2013probing}. \nick{However, as the 2-torus is not homeomorphic to the torus, \tom{due to the differing topology}, one might expect the system dynamics within each section of the double-ring to be different from the single-ring case}. We
have verified that approximating the single-ring potential as $V_{\text{ext}%
}\propto (\rho -R)^{2}$, instead of the expression adopted in Eq.~(\ref{eqn:dbpot}), does not conspicuously affect results presented below. Panel
(i) in Fig.~\ref{fig:dr}(a) shows potential \eqref{eqn:dbpot} for $R=25~%
\mathrm{\mu }$m, $w=6~\mathrm{\mu }$m and $\delta =0$, and the blue curve in
(ii) shows a cross section along $y=0$. These trap parameters are chosen to
fit known experimental ranges, typically $R=(12$--$70)\,~\mathrm{\mu }$m and
{$w=(3$--$12)\,$}$~\mathrm{\mu }$m \cite{corman2014quench,bell2016bose}.
Experiments with even larger radii, reaching $R=262\,~\mathrm{\mu }$m, have
been performed with time-averaged potentials; however the condensate was not
coherent across the whole ring in this setup \cite{sherlock2011time}.

Following the instantaneous quench, the atomic density gradually
grows to the equilibrium value, while the phase is relaxing to a
steady-state configuration. The process is random, with each numerical run
proceeding differently.~Vortices, spontaneously created in the course of the
growth of the condensate, eventually decay, potentially leaving a persistent
current with winding numbers, $n_{L}$ and $n_{R}$, in the left and right rings, respectively.

Figure \ref{fig:dr}(b) displays these dynamics, as the density equilibrates
in panels (i)-(v), and the coalescence of the phase patterns is observed in
(vi)-(x). For clarity's sake, the phase plots are masked by the
Heaviside function, $\Theta (0.9V_{0}-V_{\text{dr}}(x,y))$, set at $90\%$ of
the potential's height, as shown by the gray shaded area in panel (ii) of Fig.~\ref%
{fig:dr}(a). In this example, the final state evolves towards an equilibrium
density with about ${N\sim 2\times 10^{5}}$ atoms and with phase winding
numbers $(n_{L},n_{R})=(1,1)$, where we use the convention of positive winding numbers for clockwise circulation \cite{rooney2013persistent}. The temporal evolution of $n_L$ ($n_R$) is shown in blue (red) in Fig.~\ref{fig:dr}(a)(iii), for a single numerical run. Each simulation leads to a random observation
of winding numbers in the left and right rings, and we seek to quantify the
effect that the presence of one ring has on the other through the stochastic
distribution of these winding numbers. \tom{A movie of the time evolution for this example is provided as supplementary material to this work.}

\begin{figure}[t]
\centering
\includegraphics[width=0.75\columnwidth]{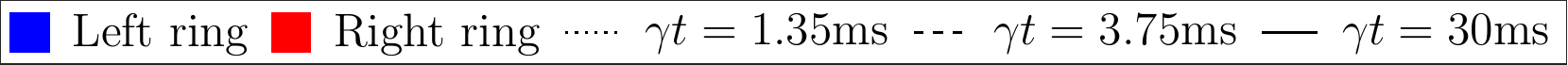}\\
\begin{minipage}{0.26\columnwidth}
\centering
\includegraphics[width=1\columnwidth]{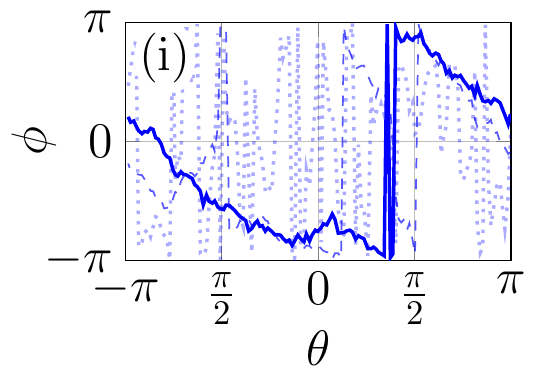}\\
\includegraphics[width=0.96\columnwidth]{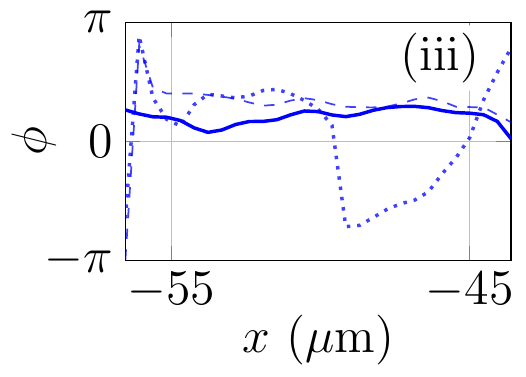}
\end{minipage}
\begin{minipage}{0.42\columnwidth}
\centering
\includegraphics[width=1\columnwidth]{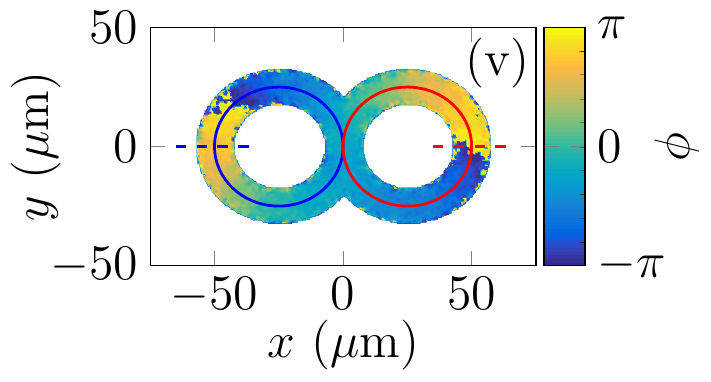}
\end{minipage}
\begin{minipage}{0.26\columnwidth}
\centering
\includegraphics[width=1\columnwidth]{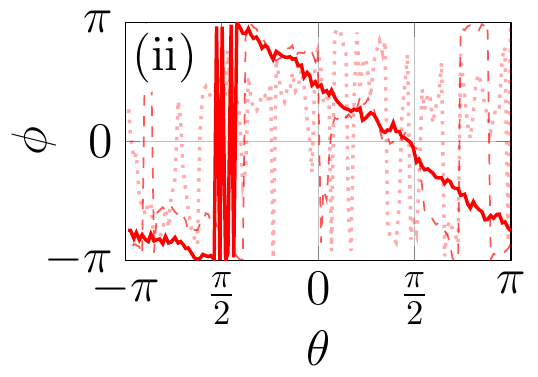}\\
\includegraphics[width=0.96\columnwidth]{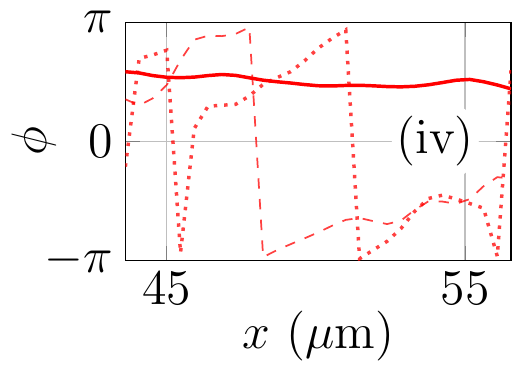}
\end{minipage}
\caption{ \tom{Azimuthal [(i) and (ii)] and radial [(iii) and (iv)] condensate phase shown for times $\gamma t=(1.35,3.75,30)$ms (dotted, dashed, solid lines respectively, with increasing opacity for later times). The azimuthal phase is extracted around the lines (i) $\rho(x+R,y)=R$ and (ii) $\rho(x-R,y)=R$, while the radial phase profile is displayed along cross section $y=0$. (v) References to the azimuthal (solid) and radial (dashed) phase profiles are superimposed on the 2D phase plot. Other parameters are the same as in Fig.~\ref{fig:dr}.}}
\label{fig:phase}
\end{figure}

Extracting the winding number from the phase pattern is a two step-process.
First, we apply high-pass filtering to the wave function in the momentum
space, setting $\tilde{\psi}(k)=0$ for $k>k_{\text{cutoff}}=\pi /\xi $, thus
removing excitations with the spatial scale smaller than a healing length, $%
\xi $, including sound waves, vortices, and thermal noise. Then, we extract the phase, $\phi (\rho ,\theta)$, from the Madelung representation of the wave function, $\psi (\rho
,\theta )=\sqrt{n(\rho ,\theta )}\exp [i\phi (\rho ,\theta )]$, at a radial
distance $\rho =R$, and count the number of jumps $\Delta \phi =2\pi $
around the ring. \tom{An example is shown in Fig.~\ref{fig:phase}(i)-(ii), where a clear $2\pi$ jump is visible in the azimuthal phase of both rings. At early times the phase is random, as shown by the dotted and dashed lines, for both the azimuthal and radial phase profiles. At later times (after the thermalisation process) a smooth radial profile appears, whilst azimuthally the phase shows evidence of a phase winding of $2\pi$ with a spatially varying gradient. The nonlinear gradient is evident in Fig.~\ref{fig:phase}(i) around $\theta=0$ and (ii) around $\theta=\pm\pi$. As we will discuss, this effect is pronounced for larger winding numbers, and a more extreme example can be found in \ref{app:phase} and in the supplemental videos provided.}

\tom{The code has been rigorously tested by manually imprinting persistent current states up to $n_w=30$ and comparing to the numerically obtained count. Of 10000 tests with varying degrees of numerical noise, none was counted incorrectly. Therefore we do not include any estimate of error in results that follow.}

Experimentally, the winding number is usually measured through a variety of
destructive techniques. In Ref.~\cite{moulder2012quantized}, a ring was
populated by two hyperfine states of $^{87}$Rb, and the interference pattern
between the rotating and non-rotating states was measured. A commonly
employed method is to measure the size of the central hole after a
time-of-flight expansion, either directly after removal of the trap \cite%
{ramanathan2011superflow,murray2013probing}, or after transforming the ring
trap into a simply connected sheet first, before turning off the trap \cite%
{murray2013probing,moulder2012quantized}. Recent advances in the application
of this technique have been achieved by the inclusion of a small stationary
BEC disk inside the annulus and measuring the interference pattern between
the disk and annulus after the expansion \cite%
{corman2014quench,aidelsburger2017relaxation}. A minimally destructive
technique has been employed to find the winding number through measuring the
Doppler shift of standing phonon modes \cite{kumar2016minimally}, allowing
for repeated measurements in the course of one experiment. There is also a
recent proposal to allow a small number of atoms tunnel into a linear
waveguide adjacent to the ring to monitor the persistent current in time \cite{safaei2019monitoring} (similar to the setup well known in optics,
with a microring \tom{resonator} coupled to a straight waveguide \cite{poon2004designing}).

\begin{figure}[!t]
\centering
\raisebox{0.25cm}{\includegraphics[width=0.415\columnwidth]{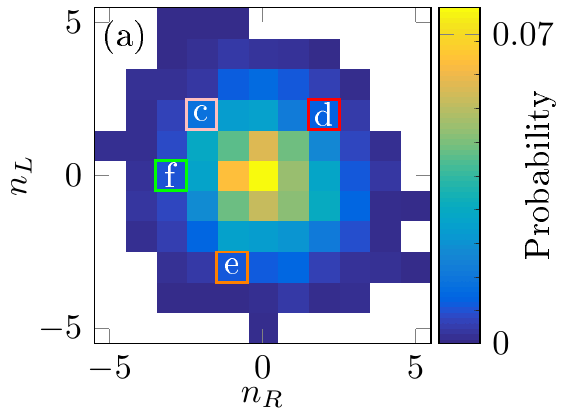}}
\includegraphics[width=0.51\columnwidth]{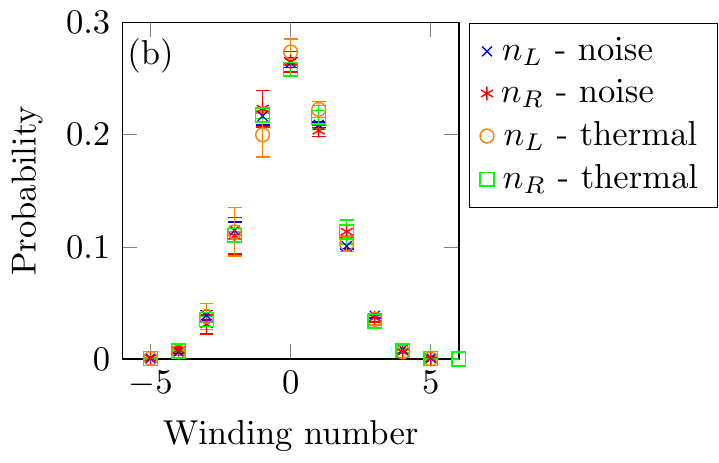}\\
\includegraphics[width=0.32\columnwidth]{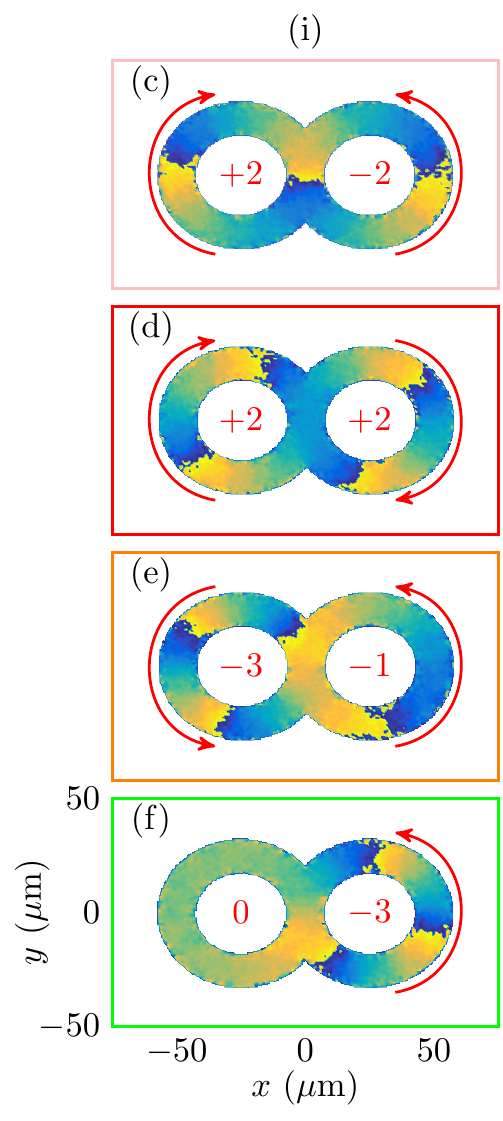} %
\includegraphics[width=0.32\columnwidth]{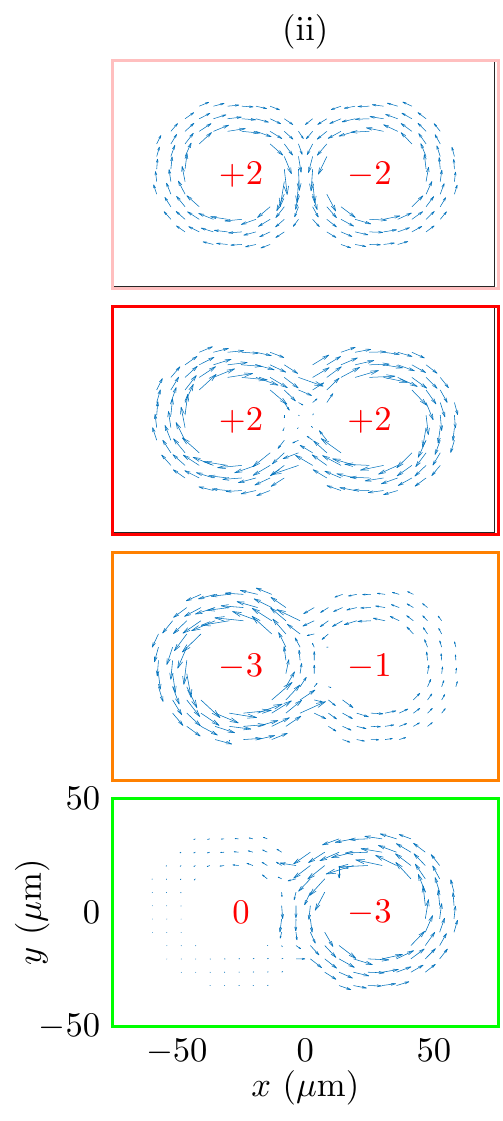}
\includegraphics[width=0.32\columnwidth]{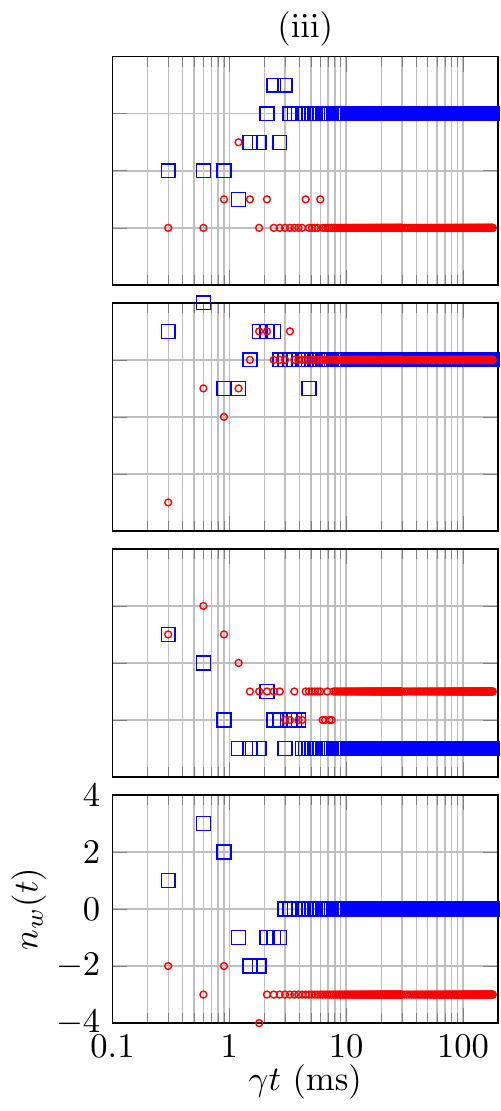}
\caption{The distribution of persistent currents in the double-ring
geometry. (a): The 2D histogram of winding numbers, produced by results of
5000 runs of the SPGPE. Coloured squares correspond to the phase
plots and velocity fields below. (b): Marginal histogram
distributions for the left ($n_{L}$) and
right ($n_{R}$) rings, comparing growth from noise to instantaneous thermal quench simulations, error bars contain the true probability within a 95\% confidence interval. (c)--(f): (i) Phase plots, (ii) velocity fields, and (iii) winding number evolutions of a typical single numerical run from within the selected squares from (a). Parameters are the same as Fig.~\protect
\ref{fig:dr}.}
\label{fig:drhist}
\end{figure}

We summarize results of the simulations by means of a 2D histogram of
distributions of winding numbers for the double-ring geometry. To
approximate the true winding number distribution, we have performed 5000
numerical runs, 
each with random initial noise. Figure~\ref{fig:drhist}(a) displays the
result, which shows the probability of observing states with winding-number
sets $(n_{L},n_{R})$, white areas indicating states that were not produced
by the simulations. \tom{The plots in (c)-(f) are examples of simulations contributing to particular points indicated on the histogram in (a). Highlighted are the states with winding-number sets ${(n_{L},n_{R})=%
\{(2,-2),(2,2),(-3,-1),(0,-3)\}}$.} This histogram displays the supercurrent pairs measured at scaled time ${\gamma t=30}$ ms; in fact, panels (iii) in Figs.~\ref{fig:drhist}(c)-(f) show that $(n_L, n_R)$ stay constant at $\gamma t \gtrsim 10$ms. The distribution of these winding numbers fits a
bivariate normal form with \emph{no correlation} between the left and right
rings: this is seen through calculation of \tom{the sample Pearson correlation coefficient}\footnote{\tom{The sample Pearson correlation coefficient is defined as
\begin{align*}
r = \frac{\sum_{i=j}^k(\text{P}(n_{L}=i)-\bar{n}_L)(\text{P}(n_{R}=i)-\bar{n}_R)}{\sqrt{\sum_{i=j}^k(\text{P}(n_{L}=i)-\bar{n}_L)^2}\sqrt{\sum_{i=j}^k(\text{P}(n_{R}=i)-\bar{n}_R)^2}}\,,
\end{align*}
where $\text{P}(n_{L}=i)$ is the probability of observing state $n_L=i$, $\bar{n}$ denotes the mean probability, and the the summation, with respect to index $i$, is performed from $j=\min(n_L,n_R)$ to $k=\max(n_L,n_R)$.}} $-1<r<1$ with $r=0$ corresponding to an uncorrelated state, with our numerical data we find $r\sim10^{-16}$.

Despite the lack of correlation in the formation of persistent currents, an effect of the neighboring ring on the other one is still visible in velocity fields ${v(x,y)=(\hbar /m)\nabla \phi (x,y)}$, seen in Fig.~\ref{fig:drhist}(ii) subplots. When the inter-ring shift in Eq.~(\ref{delta}) is $\delta =0$, the speed of
atoms across the overlap region, where the two ring traps are abutting on
each other, may be approximated by the relation
\begin{equation}
|v(0,0)|=\frac{\hbar }{m}\frac{|n_{L}-n_{R}|}{R}\,.
\end{equation}%
The validity of this relation is most easily observed in panel (ii) Fig.~%
\ref{fig:drhist}(d), with ${(n_{L},n_{R})=(2,2)}$. The speed is clearly zero
at the centre due to the counter-flow between the two rings; nevertheless,
the total phase winding around each ring is still $4\pi $. We conclude that
when, $n_{L}=n_{R}$, atoms are quiescent at the centre due to the
counter-flow. The atom fluxes flow along outside path, still maintaining the
total phase winding around each ring.

However, when $n_{L}=-n_{R}$, there is a co-flow across the centre, with the
speed doubled in the central reservoir. This scenario is shown in panel (ii)
of Fig.~\ref{fig:drhist}(c), where the velocity for the case of $%
(n_{L},n_{R})=(2,-2)$ is shown as a function of the angle around the ring.
The velocity is indeed doubled across the overlap of the two rings, and
correspondingly suppressed around the rest of the rings, such that values $%
2\pi n_{w}$ of the total phase windings are maintained in the system.

The atom flow for all other cases, with $|n_{L}|\neq |n_{R}|$, leads to an
exchange of atoms between the rings, determined by the magnitude of $%
|n_{L}-n_{R}|$. Perhaps the most interesting case is displayed in panel (ii)
of Fig.~\ref{fig:drhist}(f), where the supercurrent in the left ring remains
zero, despite constantly exchanging atoms with the right counterpart. This
is evidenced by the phase gradients present around the left ring, while the
maintained total phase accumulation is zero.

\tom{The impact of these results is apparent when considering potential experimental measurements of the current. Destructive measurements are unlikely to be possible, given the close proximity of the two rings. However, the minimally destructive measurement techniques, currently applied to single-ring geometries, will be affected by the angular dependence of the velocity around each ring, shown in Fig.~\ref{fig:drhist}(ii). This will induce a spatial dependence on the  Doppler shift of phonon modes \cite{kumar2016minimally} and the atom flux entering an adjacent linear waveguide \cite{safaei2019monitoring}.}


\subsection{Comparison to instantaneous thermal quenches}
The use of a random noise initial condition is a numerically less demanding approximation to computing an instantaneous thermal quench. To justify our choice for the former, we compare the distribution obtained through a dynamical quench to that of one obtained from a random initial state. Performing a thermal quench simulation requires equilibrating to a thermal cloud with temperature $T=300$nK, then instantaneously quenching to the target temperature $T=10$nK at $t=0$. As before, the winding numbers are measured at $\gamma t=30$ms. Summing over the observed probabilities with fixed
$n_{L}$ (or $n_{R}$) from the histogram yields a marginal
distribution for $n_{R}$ (or $n_{L}$). Figure~\ref{fig:drhist}(b) shows that both methods produce practically identical results, and thus we choose to equilibrate from noise for the rest of this work. Error bars represent a 95\% confidence interval containing the true probability, found by fitting with a Normal distribution.

\section{Benchmarking against instantaneous quenches in a single ring}

\label{sec:srr}

The observation of the formation of uncorrelated persistent currents in the
two rings in both \tom{instantaneous quenches from noise or from a thermal initial state, despite the evident density overlap}, suggests that such a distribution may \nick{in fact} be explained in terms of well-known results for independent single-ring settings. 
\nick{In quenches from a noisy initial condition, this result might have been anticipated by extrapolation of the 1D findings of Das {\em et al.}~\cite{das2012winding}, who highlighted the critical importance of {\em local} phase formation and evolution, over the global evolution around the ring. However in the present case, this was by no means {\em a priori} guaranteed for two reasons: firstly, the simulations in Ref.~\cite{das2012winding} were limited to an explicitly 1D ring, as opposed to the 2D rings numerically simulated here, which also account for the role of radial phase fluctuations (see Fig.~\ref{fig:phase} and movies); secondly, and most importantly, given that the 2-torus is not homeomorphic to the torus, the final results in the two cases need not be mappable onto each other. Indeed, while the overall distribution of winding numbers remains the same across the two non-homeomorphic cases, the actual azimuthal phase gradient is non-uniform in the case of the double ring, in stark contrast to the uniform phase gradient of a single ring.
In the end, our explicit 2D numerical simulations confirm that the local character of phase evolution also dominates in the 2D case in the determination of the {\em long-term} winding number combinations across the two rings, while the short-term evolution is random.}

\begin{figure}[tbp]
\raisebox{2.2cm}{
\begin{minipage}{0.25\textwidth}
\includegraphics[width=1\columnwidth]{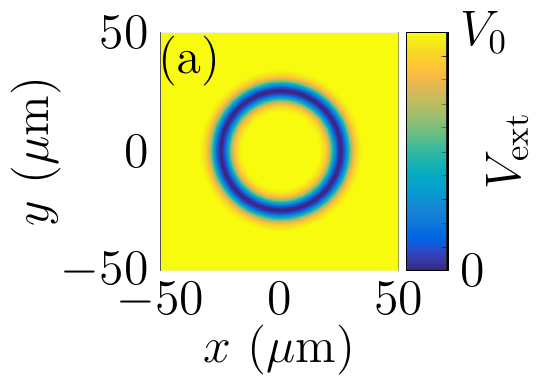}\\
\includegraphics[width=1\columnwidth]{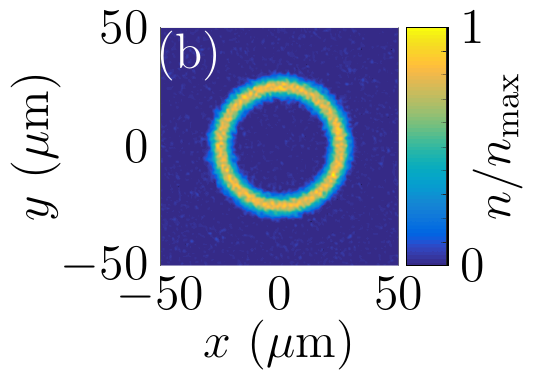}
\end{minipage}
} \raisebox{0.0cm}{\includegraphics[width=0.4%
\columnwidth]{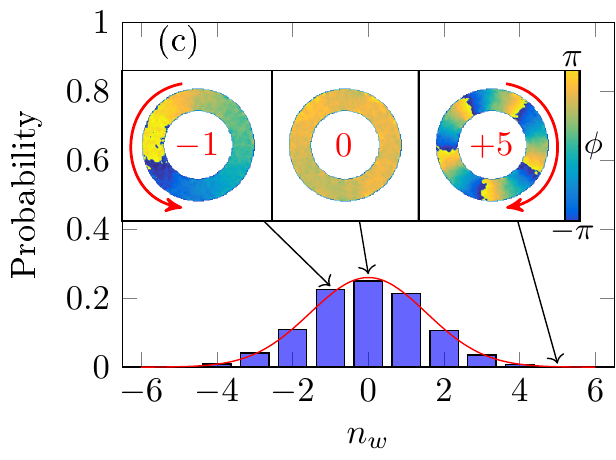}} \includegraphics[width=0.3%
\columnwidth]{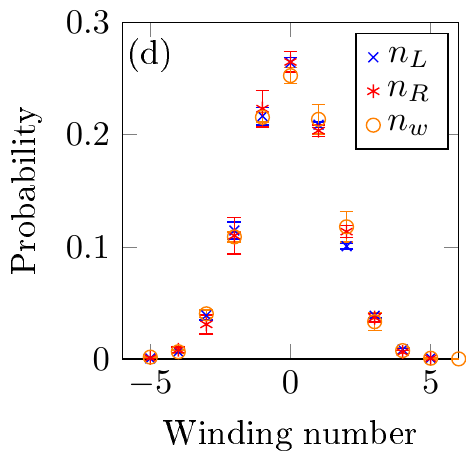}
\caption{Formation of persistent currents in the single-ring trap. (a) The
trapping potential from Eq.~\eqref{eqn:srpot} with parameters $R=25~\mathrm{%
\protect\mu }$m and $w=6\ \mathrm{\protect\mu }$m. (b) An equilibrium density profile
produced by a single run of the SPGPE. (c) The distribution
of winding numbers $n_{w}$ after $5000$ realisations of the SPGPE, shown
along with a fitted Gaussian distribution (red curve), whose standard
deviation is $\protect\sigma =1.5338$. Insets show examples of phase
profiles for $n_{w}=(-1,0,5)$ from left to right, masked by the Heaviside
function $\Theta (0.9V_{0}-V_{\text{sr}}(x,y))$ for clarity. (d) Marginal
distributions for the single-ring $n_{w}$, and for the left ($n_{L}$) and
right ($n_{R}$) rings from Fig.~\protect\ref{fig:drhist}.}
\label{fig:sr}
\end{figure}

To make a connection to the single-ring case, we consider a single ring, defined by potential
\begin{equation}
V_{\text{sr}}(x,y)=V_{0}\left\{ 1-\exp \left[ -2\left( \rho (x,y)-R\right)
^{2}/w^{2}\right] \right\} \,.
\label{eqn:srpot}
\end{equation}%
For ease of comparison, we keep here all parameters the same as in the
previous section, and observe the distribution of winding numbers after an
instantaneous quench in such a ring. In particular, for $\mu $ kept fixed,
the arising atom number in the single trap is found to be $N\sim 1.1\times
10^{5}$. Our findings are presented in Fig.~\ref{fig:sr}, showing explicitly
the potential [panel~\ref{fig:sr}(a)], and the equilibrium density at $%
\gamma t=30~$ms in panel (b).

The obtained distribution of winding numbers, $n_{w}$, after 5000 numerical
runs [Fig.~\ref{fig:sr}(c)] reveals a (univariate) normal distribution $%
N_{w}\sim N(0,\sigma ^{2})$, with ${\sigma =1.5338\pm 0.037}$ within a 95\%
confidence interval; this is plotted as a red curve in the histogram of Fig.~\ref{fig:sr}(c).
There are several factors that can reduce the observed value of $\sigma $,
including a trap's tilt angle away from the azimuthal plane \cite%
{das2012winding}, a longer thermal quench time \cite{corman2014quench}, and
smaller chemical potentials \cite{aidelsburger2017relaxation}.

Zurek's original work considered how the thermal quench through a phase transition would leave behind a superfluid circulation in an annulus \cite{zurek1985cosmological}.
In the course of the phase transition, the condensate forms $N\approx C/d$
uniformly spaced, independent regions of coherent phase around the ring,
with circumference $C$ and defect size $d$. Taking $d\sim w$ \cite%
{zurek1985cosmological} gives $N=2\pi R/w$ regions. Paraoanu derived~\cite%
{paraoanu2003persistent} that for $N$ independent condensates with uniform
phases, placed alongside one another, the maximum stable winding number is $%
N/4$. Thus, for these parameters we would expect winding numbers $n_{w}<7$,
which is consistent with our finding of $n_{w}=6$ being the largest recorded
winding number.

Further confidence in our methodology and presented numerical predictions is
provided by explicitly comparing our numerical results to findings of a
recent experiment with ultracold atoms on a \nick{ single ring.}
 Specifically, Corman {\em et al.} \cite{corman2014quench} considered finite-duration quenches on a quasi-2D ring,
in a direct test of the Kibble-Zurek scaling law. Simulating \nick{such a finite-duration cooling quench protocol
by means of the present scheme and starting from an initial thermal state}, we obtain excellent agreement (see
\ref{app:comp}).

To compare the results obtained in the framework of the double-ring
geometry, to those considered in the single-ring geometry, we compare to the marginal distributions from Fig.~\ref{fig:drhist}. These results are practically identical to the
distribution of the winding numbers for the single ring, as clearly seen in
Fig.~\ref{fig:sr}(d). We thus conclude that a robust property, inherent to
each ring, is that, under quenched spontaneous condensate growth dynamics, the
 presence of the second ring does not alter the steady-state distributions of persistent currents in a given one, \nick{due to the prevailing importance of {\em local} phase formation and evolution even in a purely 2D setting. Despite the non-homeomorphic nature of the potentials, the close agreement between
the final `steady-state' results for the two rings and the single one is understood by the fact that
the current is a topologically protected number associated with each ring}
(it may be thought of as having a `ghost' vortex in the centre of the ring).%

\section{Dependence on the trap geometry}

\label{sec:trap}

\begin{figure}[tbp]
\centering
\begin{minipage}{0.49\columnwidth}
\centering
\hspace{1.5cm}\includegraphics[width=0.8\columnwidth]{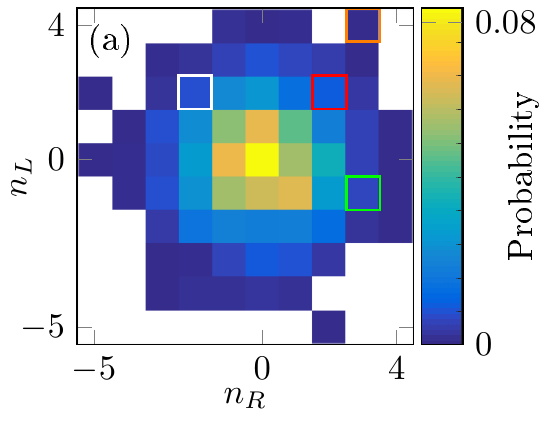}\\
\includegraphics[width=0.75\columnwidth]{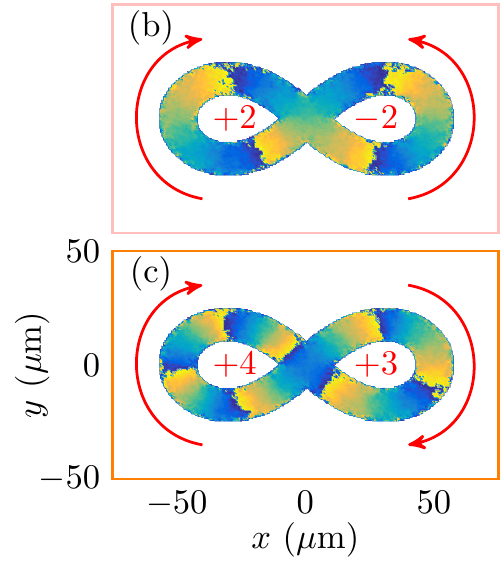}
\end{minipage}
\begin{minipage}{0.49\columnwidth}
\centering
\includegraphics[width=0.75\columnwidth]{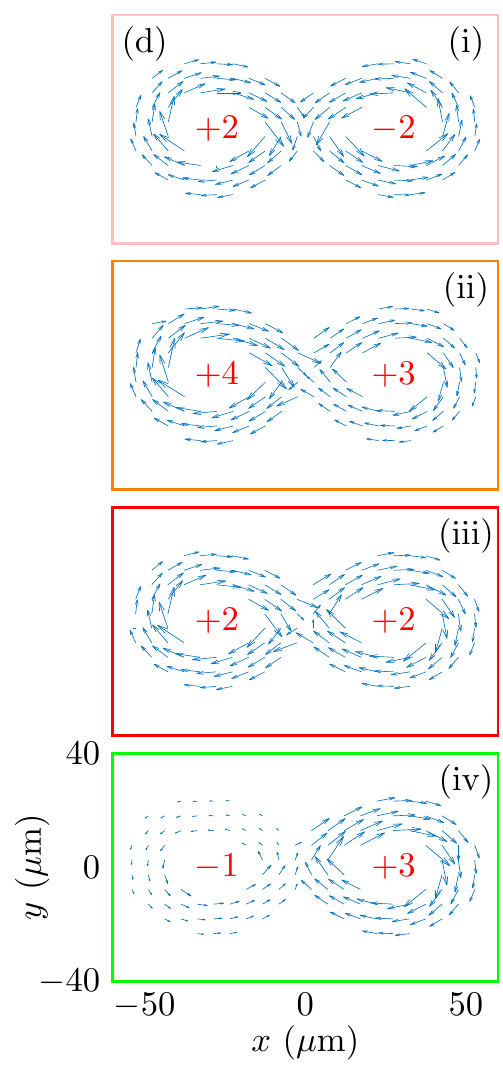}
\end{minipage}
\caption{The quench in the 2D Bose gas confined in the lemniscate potential.
(a): The 2D histogram of winding numbers produced by $5000$ runs of
simulations of the SPGPE. Coloured squares correspond to the surrounding
phase plots and velocity fields. (b)-(c): Selected phase plots, showing
winding-number sets $(2,-2)$ and $(4,3)$. (d): Panels (i)-(iv) display
velocity fields of selected squares from (a). Other parameters are the same
as in Fig.~\protect\ref{fig:dr}.}
\label{fig:lemn}
\end{figure}

The robustness of the winding number can also be tested by replacing
the two identical rings by
the Bernoulli lemniscate, or figure-of-eight configuration \cite{dalibard}. It is defined by the in-plane potential%
%
\begin{equation}
V_{\text{lem}}(x,y)=V_{0}\left( 1-\exp \left\{ -2\left( \min_t \left[ \left(
x-x_{\text{lem}}(t)\right) ^{2}\right] +\min_t \left[ \left( y-y_{\text{lem}%
}(t)\right) ^{2}\right] \right) /w^{2}\right\} \right) \,,
\end{equation}%
where $x_{\text{lem}}$ and $y_{\text{lem}}$ are described parametrically for
$t\in (0,2\pi )$ as
\begin{equation}
x_{\text{lem}}(t)=\frac{2R\cos t}{\sin ^{2}t+1}\,,\hspace{1cm}\mathrm{and}%
\hspace{1cm}y_{\text{lem}}(t)=\frac{2R\cos t\sin t}{\sin ^{2}t+1}\,.
\end{equation}%
The choice of an effective radius $R=a/\sqrt{2}$, where $a
$ is the length from the origin to the foci, gives two rings of comparable radius to the work above.

The results corresponding to this geometry are shown in Fig.~\ref%
{fig:lemn}. Although the overlap area between the two ring-like structures
is significantly different in this case, with flow from each ring structure
in the central region directly passing through each other, we still observe that the
winding numbers in each ring are completely uncorrelated. Although, in this
geometry, the velocity fields show exchange of atoms between the two
circulation loops, there is no transfer of persistent currents between them.

We also considered the case of a finite shift between the two coupled ring
traps, i.e., $\delta \neq 0$ in Eq.~(\ref{delta}). In the case of co-flows ($%
n_{L}=-n_{R}$), patterns of the fluid flow are similar to those in the case
of $\delta =0$. However, for realizations exhibiting counter-flows ($%
n_{L}=n_{R}$) varying $\delta $ changes the interpretation of the flow
circuit. At $0<\delta <w$, the shear flow between the rings can create a
vortex in the low-density overlap region, or several vortices, if the
winding number is large enough. In this scenario, the Kelvin-Helmholtz
instability may develop at the interface \cite{baggaley2018kelvin}, as seen
in simulations of the merger of two stacked rings with circulation \cite%
{oliinyk2020nonlinear,oliinyk2019symmetry}, see also \cite{chen2019immiscible}. An example of this for $\delta=w/2$ is shown in \ref{app:del}.

All cases considered above uphold our conclusion that, in our geometry, the
resulting winding numbers in both rings are produced by the random phase
profiles spontaneously emerging in the course of the condensate growth, and
the topological stability of the established states is maintained by the
presence of persistent currents in each closed ring-shaped geometry.


\section{Discussion}

\label{sec:conc}

We have explored the spontaneous growth of persistent currents in a quenched
2D Bose gas in the co-planar side-by-side double-ring geometry. 
The emerging persistent
currents are stable and long-lived, and do not transfer between the rings.
The \nick{overall value of the winding number in each ring behaves independently, which
can be directly attributed to the importance of the local nature of phase establishment in the azimuthal direction, an observation that remains valid even when allowing for radial random phase variations. While the distribution of winding numbers across the two rings can therefore be directly mapped onto the well-known results for the single-ring setting, it is important to note that the azimuthal phase gradient in each ring (for a given non-zero winding number) is {\em not} constant, in direct contrast to the constant phase gradient of the single ring case. This is more clearly visible in cases of large $|n_L|+|n_R|$,  examples of which are shown in \ref{app:phase} and movies of the real-time evolution of Figs.~\ref{fig:dr} and~\ref{fig:nonlinphase}. This is one manifestation of the non-homeomorphism between a single torus and a 2-torus. Remarkably, however, this is not necessarily an impediment to potential future atomtronic devices, which could utilize the observed robust nature of the superfluid current configuration. The independence of the ring winding numbers and stability of the formed supercurrents may facilitate
 a well-controlled current transfer protocol between rings. Such deterministic transfer of winding numbers across multiple-loop atomtronic architectures is a promising direction for future research.}

Varying the separation of the rings demonstrates the ability to modify the
atom transfer between the rings and generate vortices between them. However,
there is no sign of angular momentum transfer between the rings. The robustness of the persistent currents after the quench holds
steadfast even against the change of trap geometry, such as replacement of
the double ring by the lemniscate potential, where one might na\"{\i}vely
expect uncorrelated persistent currents in individual rings to be suppressed due to
atomic motion along a\textquotedblleft figure-of-eight" path. We also varied
the phenomenological damping parameter $\gamma $ in a broad range of values,
and extended the simulation time, which produced no evidence of decay,
confirming the efficiency of the topological protection of the spontaneously
generated winding numbers.%

A natural question suggested by the present results is the transfer of the
winding numbers between the rings, which will be the subject of further
work. The ability to control the transfer of winding numbers could be
envisaged as a prototypical atomtronic switch, providing an avenue for
controllable realization of coherent quantum phase slips required for the
Mooij-Harmans qubit \cite{mooij2005phase,mooij2006superconducting}. A hot
topic of current studies of toroidal BECs is the supercurrent decay
mechanism, an understanding of which could help one to control the transfer
of current states between the rings. The Gross-Pitaevskii equation and its
many extensions \cite{blakie2008dynamics,proukakis2008finite,Proukakis13} do not capture supercurrent decay, even at finite temperature,
without an external barrier. However, due to roughness of the potential, the
decay time for large winding numbers decay in experiments is on the order of
seconds, whereas $n_{w}\sim 1$ may be stable on times on the order of
minutes \cite{moulder2012quantized}. Theoretical studies of the decay
mechanism so far have all relied on effects along the annulus, caused by a
repulsive barrier, the decay being visualized as vortices crossing the
barrier region radially. In Ref.~\cite{kumar2017temperature} it was shown
that the temperature-induced decay does not fit the Caldeira-Leggett
superconductivity model, i.e.,~the observed decay rate does not match simple
models of quantum tunnelling and thermal activation of phase slips. Using a
truncated Wigner approximation, Ref.~\cite{mathey2014decay} attributed some
of the disagreements between the theory and experiment to thermal
fluctuations, however exact identification of the decay mechanism remains an
open question.

Data supporting this publication is openly available under
an Open Data Commons Open Database License \cite{data}.

\section{Acknowledgements}

We would like to thank Fabrizio Larcher for his early involvement in this
work. We thank I-Kang Liu, George Stagg and Jean Dalibard for discussions and J\'{e}r\^{o}me Beugnon for comments on the manuscript and providing us with the experimental data of Ref.~\cite{corman2014quench}. TB thanks EPSRC Doctoral Prize Fellowship Grant No.~EP/R51309X/1. We acknowledge financial support from the Quantera ERA-NET cofund
project NAQUAS through the Engineering and Physical Science Research
Council, Grant No.~EP/R043434/1. The work of BAM who also acknowledges a Visiting Professorship held at Newcastle University is supported, in part, by
the Israel Science Foundation through grant No.~1287/17.

\appendix

\section{Nonlinear azimuthal phase gradient around the ring}\label{app:phase}

\begin{figure}[t]
\centering
\includegraphics[width=0.28\columnwidth]{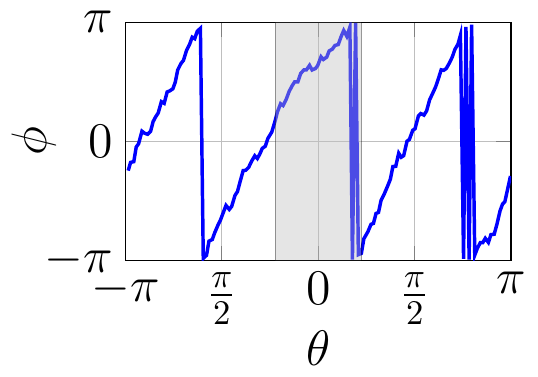}
\includegraphics[width=0.36\columnwidth]{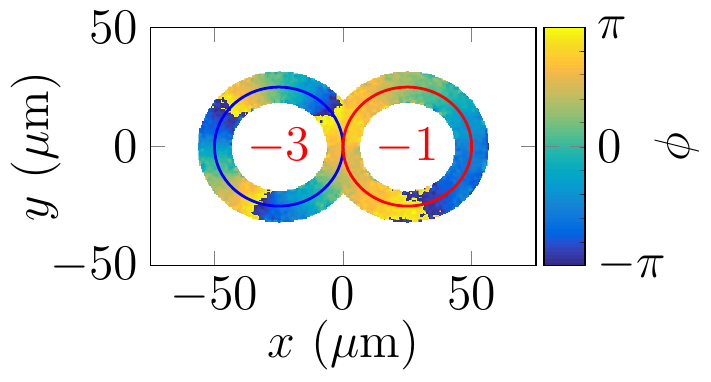}
\includegraphics[width=0.28\columnwidth]{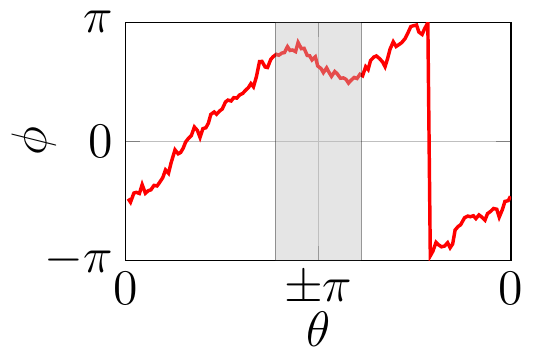}
\caption{\tom{Centre: 2D phase profile for the case $(-3,-1)$. Left: Azimuthal phase profile for the left ring, grey highlights the region of overlap between the rings. Right: Azimuthal phase for the right ring, shifted in the $x$-axis such that the overlap region of the two rings occurs in the grey region. Other parameters are the same as in Fig.~\ref{fig:dr}.} }
\label{fig:nonlinphase}
\end{figure}

\tom{We have shown that in co-planar two-ring geometries with large density overlap the phase gradient for a persistent current winding is not linear, but instead a function of the azimuthal angle $\theta$ around the ring. In Fig.~\ref{fig:nonlinphase} we illustrate this effect for the supercurrent combination $(-3,-1)$. Crucially, the presence of the flow from the left ring is strong enough to {\em reverse} the direction of the flow in the right ring, as evidenced from the grey-shaded region pf overlap of the two rings. As explained in Sec.~\ref{sec:drr}, this is possible due to the exchange of atoms between the rings. The time evolution of the density and phase (shown in the 2D plane, azimuthally and radially) can be seen in the supplemental material provided with this work.
}

\section{Comparison of experimental results and theoretical predictions}\label{app:comp}

In 1985 Zurek considered how a thermal quench through a phase
transition would leave behind a superfluid circulation in an annulus \cite{zurek1985cosmological}. He postulated that during the phase transition the condensate forms $N\approx C/d$ independent regions of coherent phase around the ring, with circumference $C$ and defect size $d$. In previous works, the resulting average winding number, calculated at the end of the thermal quench, has been shown to scale as
\begin{equation}
\expval{|n_w|}\propto \left( \frac{C}{d}\right) ^{1/2}\sim \left( \frac{2\pi
R}{w}\right) ^{1/2}\,,
\label{eqn:abswind}
\end{equation}%
where we assume that the width of the ring is close to the defect's size, $%
d\sim w$ \cite{zurek1985cosmological}.

\subsection{Direct confirmation of Kibble-Zurek scaling for finite-duration quenches}
\begin{figure}[tbp]
\centering
\includegraphics[width=0.48\columnwidth]{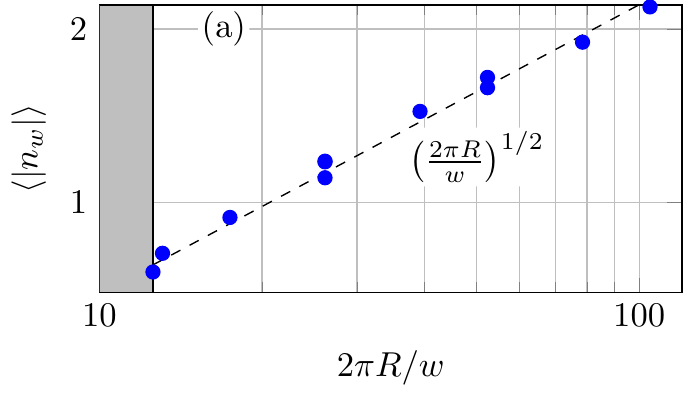} \raisebox{0.2cm}{%
\includegraphics[width=0.48\columnwidth]{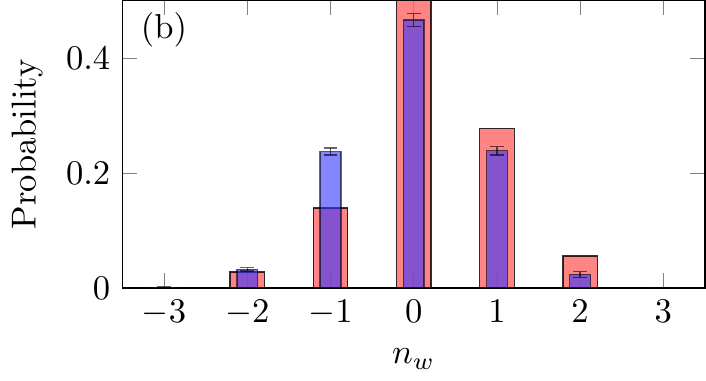}}
\caption{(a) A power-law dependence given by Eq.~\eqref{eqn:abswind}. \tom{Each blue point is the measurement of $\expval{|n_w|}$ after 5000 SPGPE simulations for varying $w=(3,6)~\mathrm{\mu }$m and $R=(12.5,18.75,25,37.5,50)\mu$m.} The gray region covers the range of $R<2w$, when there is no central hole in the density. (b) The distribution of observed winding numbers, produced by the
thermal quench with ramp time $\tau_{r}=0.025$s. Red (wide) bars are experimental data from Ref.~\protect\cite{corman2014quench} for the single ring, based on 36 runs, while blue (thin) bars represent numerical data collected from 5000 realisations of the SPGPE. Parameters are specified in the main text.}
\label{fig:comp}
\end{figure}

\tom{In the related work of Das {\em et al.} \cite{das2012winding} they tested the dependence of relation \eqref{eqn:abswind} on a linear temperature quench in time in an explicit 1D setting, and found excellent agreement. In this Appendix we carry out finite duration quenches from a high to a low temperature, in an explicitly 2D geometry, thus assessing the 2D nature through the radial width $w$. In Fig.~\ref{fig:comp}(a) it is seen that our analysis agrees with this prediction, indicating optimal experimental geometries for observing larger winding numbers. Specifically here we have chosen $w=(3,6)~\mathrm{\mu }$m$\,\gg\xi$ and varied $R=(12.5,18.75,25,37.5,50)\mu$m using a finite duration quench with the ramp time $\tau_r=0.001$s from an initial thermal state at $T=250~$nK ($T\sim 1.25T^\infty_\text{BKT}$) to $T=10$nK (${T\sim 0.05T^\infty_\text{BKT}}$)}.




\subsection{Comparison to experiment of Ref.~\cite{corman2014quench}}
Next we compare our results to the recent work by Corman \textit{et al.} \cite{corman2014quench}%
, which addressed temperature quenches with different ramp rates in the
single-ring geometry with parameters $R=12~\mathrm{\mu }$m and $w=3~\mathrm{%
\mu }$m. We take ${\mu =12.5k_{\mathrm{B}}~}%
\mathrm{nK}$ to match the atom number $N=36000$. At the end of each temperature quench, the winding number was
measured by turning off the trap potential and looking at the ensuing
interference pattern, produced by the interplay of the ring with internal
stationary disk. We have carried out simulations of one of those
experiments, for the finite duration thermal quench from $T=300~$nK ($T\sim 2T^\infty_\text{BKT}$) down to $T=10$ nK ($T\sim 0.07T^\infty_\text{BKT}$), with
the ramp time $\tau_r=0.025$s. The experiment was repeated $36$
times, with the aim to create a histogram of the resulting winding numbers.
In Fig.~\ref{fig:comp} we compare the histograms representing the
experimental findings and our numerical results  (red and blue columns,
respectively), the latter ones produced by $5000$ simulations of the SPGPE.
The figure demonstrates excellent agreement. Due to a relatively low number
of experimental realizations, the respective histogram is not exactly symmetric
about $n_{w}=0$, although it features $\expval{n_w}\approx 0$, confirming
the stochasticity of the distribution. The experimentally produced average
absolute winding number corresponding to this dataset is $\expval{|n_w|}=0.6$%
, while our simulations yield $\expval{|n_w|}=0.5926$. The experiments did
not feature winding numbers $|n_{w}|>2$, while $5$ of our $5000$ simulations
yielded $|n_{w}|=3$. It may be that still larger values are possible in this
geometry, but with a probability $<0.001$. Comparing to the relation from
Ref.~\cite{paraoanu2003persistent}, we expect that the maximum permitted
value is $|n_{w}|=6$, although this prediction does not account for the temperature ramp
rate.

\section{Effect of finite ring separation distance
$\protect\delta $ on winding number histogram}
\label{app:del}

\begin{figure}[!t]
\centering
\begin{minipage}{0.25\textwidth}
\includegraphics[width=1\columnwidth]{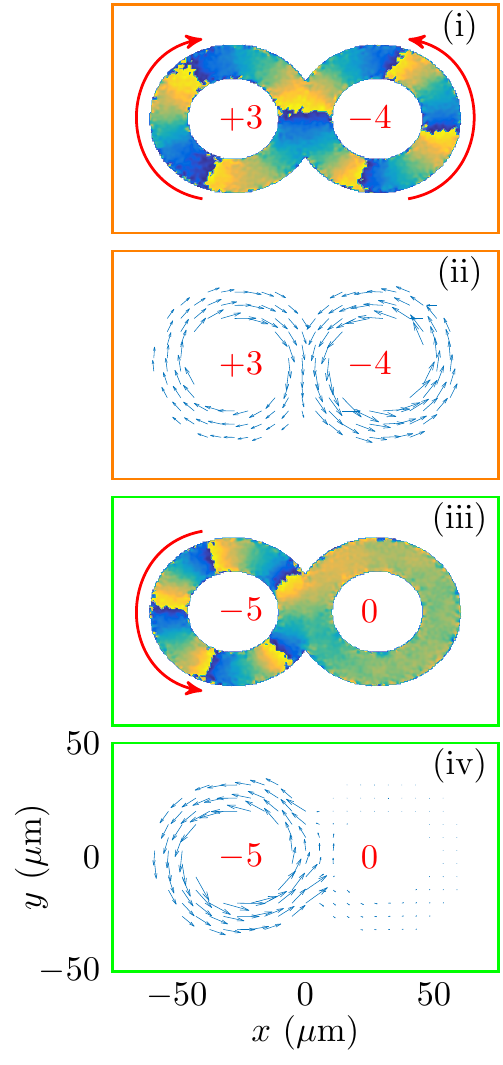}
\end{minipage}
\begin{minipage}{0.45\textwidth}
\centering
\includegraphics[width=0.8\columnwidth]{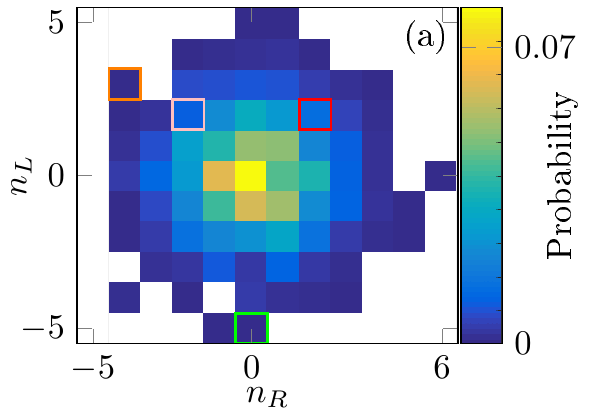} \\
\includegraphics[width=0.8\columnwidth]{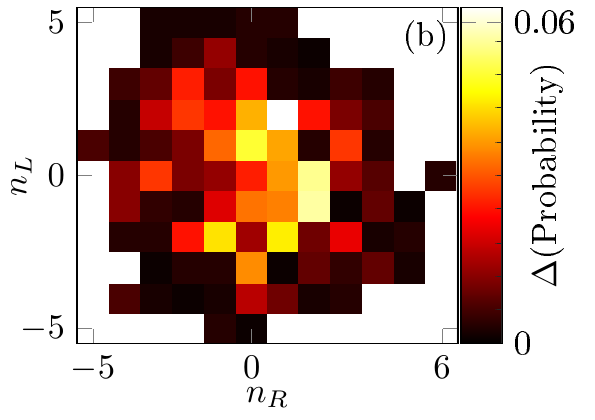}
\end{minipage}
\begin{minipage}{0.25\textwidth}
\includegraphics[width=1\columnwidth]{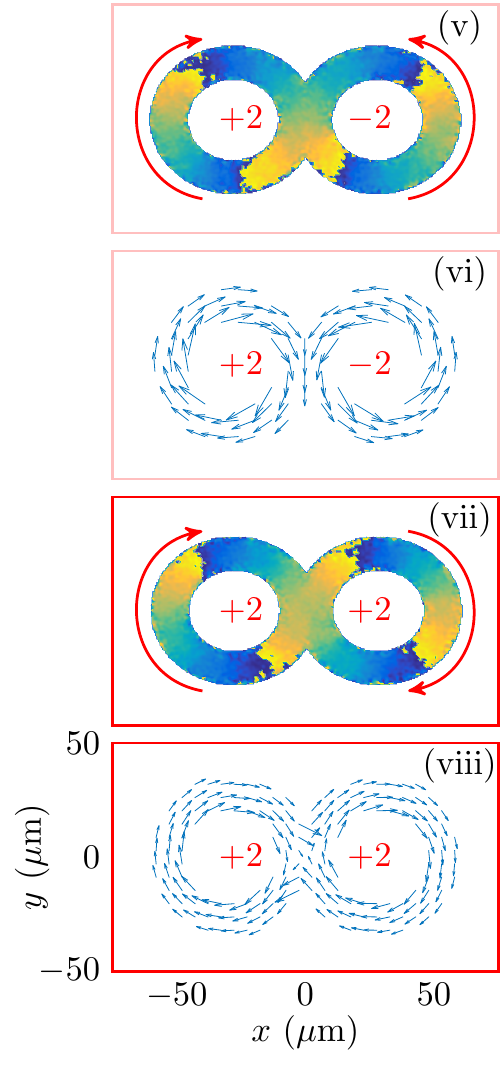}
\end{minipage}
\caption{Distribution of persistent currents in a double-ring geometry, with
$\protect\delta=w/2$. (a) 2D histogram of observed winding numbers after
5000 runs of the SPGPE. Coloured squares correspond to the surrounding phase
plots and velocity fields. (b) Measure of the relative difference between the $\protect%
\delta=0$ and $\protect\delta=w/2$ histograms. Note the different spatial extent of the colorbar in this case, corresponding to a maximum relative error of 6.3\%. (i)--(viii) Phase plots and
velocity fields of selected squares from (a). Other parameters are the same
as Fig.~\protect\ref{fig:dr}.}
\label{fig:drhistdel}
\end{figure}

Addressing effects of the finite separation between the rings [$\delta >0$
in Eq.~(\ref{delta})], in Fig.~\ref{fig:drhistdel} we present results of $%
5000$ simulations of the SPGPE with $\delta =w/2$. The histogram of steady-state winding numbers does not
display any significant difference from the $\delta =0$ case. This conclusion was
verified by calculating the relative error between the two observed \tom{probability}
distributions, \tom{${\Delta \text{(Probability)}=|\text{P}(n_{w}(\delta =0))-\text{P}(n_{w}(\delta=w/2))|/\text{P}(n_{w}(\delta =0))}$}, the result being that the variation is $<7\%$. The correlation of the winding numbers between the two separated rings is still
effectively zero ($r\sim10^{-15}$). Note that, in the limit of large $%
\delta $, the rings become completely independent systems, for which the
above results for the single ring are directly relevant.

For intermediate values of $\delta$ ($0<\delta<w$) the velocity fields are remarkably similar to those presented for $\delta=0$.
 However, as stated in the main text, states with $n_{L}=n_{R}$ exhibit shear flow between the rings which may create vortices in the low density overlap region \cite{baggaley2018kelvin,oliinyk2020nonlinear,%
 oliinyk2019symmetry}.
 
 At $\delta >w$ the rings are spatially separated, with little density overlap. Recent works have found that, even in this case, the angular-momentum states (truncated to $|n_{w}|\leq 1$) can couple to one another and tunnel, at the single-particle level \cite%
{polo2016geometrically,pelegri2017single,pelegri2019topological,pelegri2019topological2}. However, the nonlinearity in the Gross-Pitaevskii equation couples the setting to
higher-order angular-momentum states, and destroys the simple picture. By
tuning the nonlinearity to be negligible through the Feshbach resonance \cite%
{stwalley1976stability}, it may be possible to create a superfluid state
admitting tunnelling of angular-momentum states.

\section*{References}

\bibliography{atomtronics}

\begin{thebibliography}{10}
\bibitem{dalibard}
This was suggested to us by {J}.~{D}alibard
\end{thebibliography}

\end{document}